\documentclass[twoside,twocolumn,english,aps,amsmath,amssymb,reprint,pra,superscriptaddress]{revtex4-1}
\usepackage[T1]{fontenc}
\usepackage[latin9]{inputenc}
\usepackage{color}
\usepackage{amsmath}
\usepackage{amssymb}
\usepackage{graphicx}

\makeatletter

\providecommand{\tabularnewline}{\\}

\usepackage{babel}
\usepackage{babel}

\makeatother

\usepackage{babel}
\begin{document}

\title{Resource reduction for distributed quantum information processing
using quantum multiplexed photons}

\author{Nicolò Lo Piparo}
\email{nicopale@gmail.com}

\selectlanguage{english}%

\affiliation{National Institute of Informatics, 2-1-2 Hitotsubashi, Chiyoda-ku,
Tokyo 101-8430, Japan.}

\author{Michael Hanks}

\affiliation{National Institute of Informatics, 2-1-2 Hitotsubashi, Chiyoda-ku,
Tokyo 101-8430, Japan.}

\author{Claude Gravel}

\affiliation{National Institute of Informatics, 2-1-2 Hitotsubashi, Chiyoda-ku,
Tokyo 101-8430, Japan.}

\author{Kae Nemoto}

\affiliation{National Institute of Informatics, 2-1-2 Hitotsubashi, Chiyoda-ku,
Tokyo 101-8430, Japan.}

\date{\today}

\author{William J. Munro}

\affiliation{NTT Basic Research Laboratories \& NTT Research Center for Theoretical
Quantum Physics, NTT Corporation, 3-1 Morinosato-Wakamiya, Atsugi,
Kanagawa, 243-0198, Japan.}

\affiliation{National Institute of Informatics, 2-1-2 Hitotsubashi, Chiyoda-ku,
Tokyo 101-8430, Japan.}

\date{\today}
\begin{abstract}
Distributed quantum information processing is based on the transmission
of quantum data over lossy channels between quantum processing nodes.
These nodes may be separated by a few microns or on planetary scale
distances, but transmission losses due to absorption/scattering in
the channel are the major source of error for most distributed\textcolor{black}{{}
quantum information tasks. Of course quantum error detection (QED)
/correction (QEC) techniques can be used to mitigate such effects
but error detection approaches have severe performance limitations
due to the signaling constraints between nodes and so error correction
approaches are preferable \textemdash{} assuming one has sufficient
high quality local operations. Typically, performance comparisons
between loss-mitigating codes assume one encoded qubit per photon.
However single photons can carry more than one qubit of information
and so our focus in this work is to explore whether loss-based QEC
codes utilizing quantum multiplexed photons are viable and advantageous,
especially as photon loss results in more than one qubit of information
being lost. We show that quantum multiplexing enables significant
resource reduction: in terms of the number of single photon sources
while at the same time maintaining (or even lowering) the number of
two-qubit gates required. Further, our multiplexing approach requires
only conventional optical gates already necessary for the implementation
of these codes.}
\end{abstract}


\maketitle
\textcolor{black}{There are many active approaches being pursued in
the development of quantum technologies, including those associated
with imaging and sensing \cite{QImaging2,QImaging3,QSensing}, communication
\cite{QComm,QC_Bill,QKD1,QKD02,QKD03,Quantumcomm} and computation
\cite{Qcomp1,Qcomp2,QComp_Bill,Quantum_comp2,Quantum_comp3,Quantum_comp4}.
What has become clear is that many of these will be distributed in
nature \cite{QC_Bill} and, as such, it will be essential to share
quantum information between the remote nodes, regardless of whether
those nodes are separated on the atomic or planetary scales \cite{NVKae,NVnetworks1,NVnetworks2}.
This distributed nature means we are going to require both a quantum
interface between matter \& photonic qubits and a photonic bus to
transfer such information between nodes \cite{interface}. However,
real implementations will suffer from losses, which will dramatically
affect the performance of the quantum protocols in which such devices
are being used. Mechanisms must be developed to mitigate such detrimental
effects.}

\textcolor{black}{There are quite a number of routes available to
offset loss effects, ranging from the development of lower loss fibers
to more efficient quantum information coding. The latter route is
quite appealing as it can be used with current technology and is likely
to be more compatible with our existing infrastructure. There is a
well known set of loss based quantum detection and error correction
codes that can be used in this situation. In \cite{Qmultiplexing}
they discuss a simple quantum network scenario in which the quantum
multiplexing (QMu) of photonic degrees of freedom allows one to design
a single-step combined entanglement distribution and error detection
protocol with improved entanglement generation rates, using fewer
physical (photons and quantum memories) and temporal resources. However,
their performance is still limited by the probabilistic nature of
the various quantum operations and the resulting necessary heralding
signals.}

\textcolor{black}{Quantum error correction codes (ECCs) naturally
avoid a heralding bottleneck, with example loss based codes including
the quantum parity \cite{redundancy_code}, cat \cite{Cochrane},
binomial \cite{Michael}, Reed-Solomon \cite{Reed_Sal}, surface \cite{surfacecode}
and GKP codes \cite{GKP1}. They allow us to approach the deterministic
transmission of quantum information over a lossy channel, as long
as those total losses do not exceed a certain threshold (50\% at most)
\cite{Holevo,wolf}. Typical encodings use either the polarization
or time bin degrees of freedom but are not particularly resource efficient
as they require a large number of single photons. However, single
photons have the potential to carry much more information using different
degrees of freedom (see Supplemental Material for further detail).
Hence the natural question is whether using multiple degree of freedom
is advantageous, especially as photon loss results in more than one
qubit of information being lost.}

\textcolor{black}{Here we investigate the potential of quantum multiplexing
to reduce the resources required to implement loss based error correction
codes. The creation of reliable single photon sources has proved challenging
and we therefore take as a central figure of merit the required number
of single photons. We analyze two well known ECCs, the redundant quantum
parity \cite{redundancy_code} and quantum Reed-Solomon codes \cite{Reed_Sal},
determining the number of photons and qubits required to reach a threshold
success probability with the multiplexing method. }

\textcolor{black}{}
\begin{figure*}[htb]
\begin{centering}
\textcolor{black}{\includegraphics[width=1\textwidth]{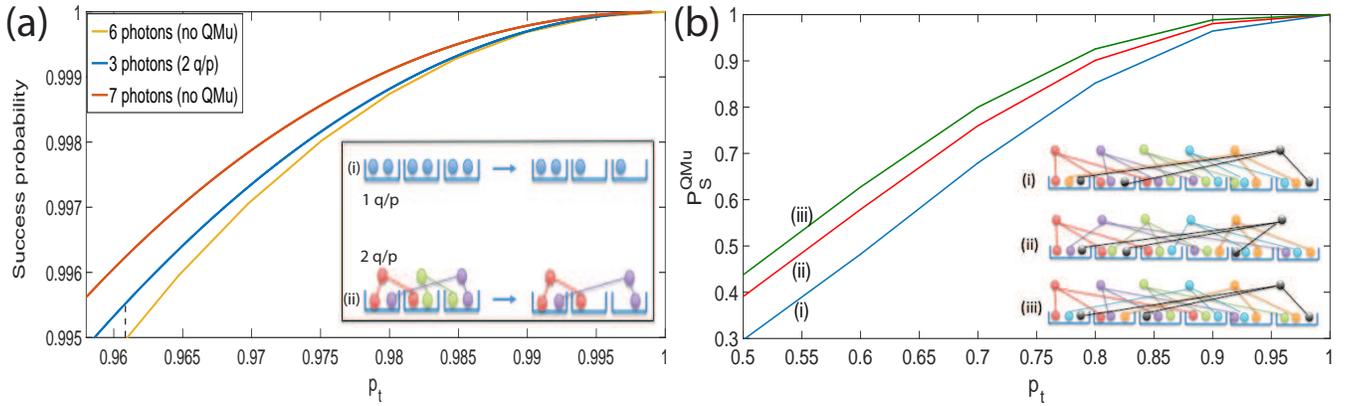}}
\par\end{centering}
\textcolor{black}{\vspace{-2em}
}

\textcolor{black}{\caption{\label{fig:1}(a) Plot of the overall success probability versus photon
transmission probability $p_{t}$ of the redundant quantum parity
code with (blue curve) and without (red and yellow curves) quantum
multiplexed photons. The inset of (a) depicts a schematic illustration
of a particular instance of the six-qubit quantum redundancy parity
code in which each photon carries one qubit (i) and three photons
carry two qubits of information each (2q/p) (ii). Similarly in (b),
we show the success probability $P_{S}^{\mathrm{QMu}}$ versus $p_{t}$
for three different configurations of a quantum multiplexed system.
Here 6 photons carry 3 qubits each, distributed over 6 blocks (each
block containing 3 qubits). }
}
\end{figure*}
\textcolor{black}{Let us begin by exploring the redundant quantum
parity code \cite{redundancy_code} in a photon transmission regime
for which both the number of qubits (memories within the node) and
the number of photons can be reduced using our quantum multiplexing
approach, all while maintaining the near deterministic transmission
of information between the two nodes. In the redundant quantum parity
code the information $\alpha,\beta$ in our encoded state $\left.|\psi\right\rangle _{\left(n,\,m\right)}=\alpha\left.|+\right\rangle _{1}...\left.|+\right\rangle _{n}+\beta\left.|-\right\rangle _{1}...\left.|-\right\rangle _{n}$
(each block term $\left.|\pm\right\rangle _{i}=\left.|H\right\rangle ^{\otimes m}\pm\left.|V\right\rangle ^{\otimes m}$
containing $m$ photons) is successfully transmitted over the channel
when at least one block of $m$ qubits arrives intact (no losses)
and each other block retains at least one photon (see Figure \ref{fig:1}a
inset). The success probability is given by \cite{Psucc_no_hyper}
\begin{eqnarray}
P_{S}=\left[1-\left(1-p_{t}\right)^{m}\right]^{n}-\left[1-p_{t}^{m}-\left(1-p_{t}\right)^{m}\right]^{n} & ,\label{PSeqn1}
\end{eqnarray}
where $p_{t}$ is the single photon transmission probability through
the channel. Our first observation is that this concatenated code
is not particularly resource efficient as the number of qubits at
the first logical layer, $m$, grows inversely with the transmission
probability $p_{t}$. Further, $n$ grows inversely with $p_{t}^{m}$
and so $(m,\,n)$ grow exponentially with distance between nodes.
Our quantum multiplexer is a natural solution \cite{Qmultiplexing}:
here we encode multiple qubits onto a single photon, meaning less
photons in total need to be transmitted. More specifically, we enact
a two qubit gate between the first degree of freedom (polarization)
and a second photonic or matter qubit. Then, swapping the polarization
of the initial photon with another degree of freedom (time bin in
this case), a third system can then interact independently with the
the polarization of this same photon (further details in Supplemental
Material). The quantum mutliplexer has many potential benefits including
deterministic operations between different degrees of freedom \textemdash{}
especially important when single photon sources are probabilistic
in nature. }

\textcolor{black}{Let us explore this in a little more detail. In
the inset of Fig. (\ref{fig:1}a.i) we illustrate a six photon redundant
quantum parity code realization without the use of quantum multiplexing
in which 3 blocks of 2 photons each are used. After the photons are
transmitted over the lossy channel, the code is successful if at least
one block contains two photons and the other two blocks each contain
one or more photons. One can think of substituting those 6 photons
with three quantum multiplexed photons each carrying two qubits of
information. In Fig. (\ref{fig:1}a.ii) these are represented by the
colored lines connected to the dots contained in the blocks (see Supplemental
Material for futher details). In this case, the ECC code only tolerates
the loss of one photon. Therefore it would seem logical that we can
reduce the number of photons by using the multiplexing approach, provided
that the success probability is above the desired threshold value.
This raises the question as to what the success probability $P_{S}^{\mathrm{QMu}}$
will be in this quantum multiplexed approach. One can show for $n_{\mathrm{tot}}$
transmitted photons that
\begin{align}
P_{S}^{\mathrm{QMu}} & =\sum_{i=0}^{n^{*}}\left[\left({n_{\mathrm{tot}}\atop i}\right)-\left(U_{i}+E_{i}\right)\right]p_{t}^{n_{\mathrm{tot}}-i}(1-p_{t})^{i},\label{eq:_general_succ_prob_hyper}
\end{align}
where $U_{i},\,E_{i}$ are the number of events in which losing $i$
photons will leave none of the blocks with the initial number of qubits
or at least one empty block and $n^{\ast}$ is the number of lost
photons the ECC code can tolerate. We need to determine both $U_{i}$
and $E_{i},$ which are highly dependent on how the quantum multiplexed
photons are connected to the blocks (see Fig. (\ref{fig:1}b)). Different
configurations lead to different success probabilities. We can also
release the constraints of all blocks having to have the same number
of qubits (an $unbalanced$ configuration), which is typically not
utilized in error correction schemes. This enables us to further reduce
the number of qubits (and photons) even in the non-multiplexed case
(see Supplemental Material).}
\begin{figure*}[t]
\begin{centering}
\textcolor{black}{\includegraphics[width=1\textwidth]{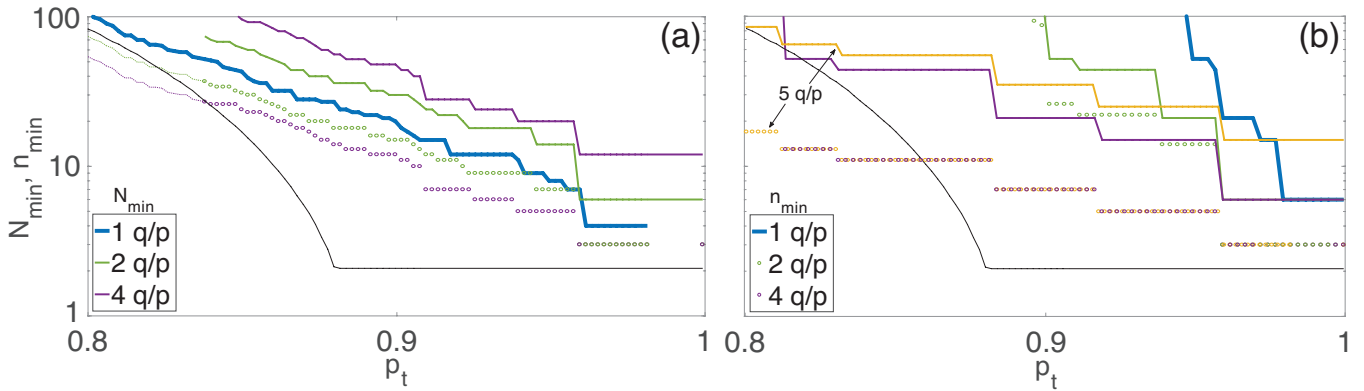}}
\par\end{centering}
\textcolor{black}{\vspace{-3em}
}

\textcolor{black}{\caption{\label{fig:qubits_photons_vs_Pt-1} Plot of the minimum number of
qubits (solid lines) and photons (dotted lines) for the (a) redundant
quantum parity and (b) Reed-Solomon codes required to reach a threshold
success probability of $\overline{P_{S}}=0.995$ versus the photon
transmission probability $p_{t}$ using a quantum multiplexed encoding
of 2 $-$ 5 qubits/photon (q/p), respectively. Also shown is the non
multiplexed situation of 1 qubit/photon (q/p) \textcolor{black}{for
all codes including the hexagonal GKP code (black curves). }}
}
\end{figure*}

\textcolor{black}{In Figure (\ref{fig:1}a) we plot the overall success
probability $P_{S}$ versus $p_{t}$ for two non-multiplexed (equal
and unbalanced) configurations alongside one quantum-multiplexed situation
with a minimum threshold success probability requirement of $\overline{P_{S}}=0.995$
(typical for many quantum computation based tasks). It is clear that
our 3 photon quantum multiplexed case (3 blocks with 2 qubits / photon)
dramatically outperforms the traditional 6 photon non-multiplexed
case (3 blocks with 2 qubits, photons each). In the region $0.958\lesssim p_{t}\lesssim0.962$
the 6 photon case does not reach our threshold target, while the 3
photon multiplexed approach does. The 7 photon configuration (with
the first block containing 3 photons while the second and third blocks
contain 2 photons each) performs slightly better than the multiplexed
case. However both are above the threshold and the multiplexed situation
uses fewer photons, qubits and two qubit gates. The multiplexed approach
also halves the number of photons in the region $0.976\lesssim p_{t}\leq0.995.$
These are critical resource savings. }

\textcolor{black}{It is clear that the lower $p_{t}$ is, the more
qubits (two qubit gates) and photons we will need to reach $\overline{P_{S}}$.
It is important, in reducing the total numbers of these resources,
to also explore unbalanced quantum multiplexing configurations. In
Figure (\ref{fig:qubits_photons_vs_Pt-1}a) we plot the minimal number
of qubits, $N_{min},$ and photons, $n_{min},$ versus $p_{t}$ for
resource-optimal configurations with 2 and 4 qubits per photon. Quantum
multiplexed systems utilize fewer photons, however the number of qubits
is either the same or slightly higher, except in a small region near
$p_{t}\sim0.97$ (Fig. (\ref{fig:1}a)). In fact we can almost halve
the number of photons being transmitted over the channel \textemdash{}
quite an advantage, especially as single photon sources are currently
not as efficient as quantum gates or measurements.}
\begin{table}[htb]
\begin{centering}
\textcolor{black}{}%
\begin{tabular}{|c|c|}
\hline 
\textcolor{black}{Total number of photons } & \textcolor{black}{Total number of qubits}\tabularnewline
\hline 
\hline 
\textcolor{black}{15 (1q/p) } & \textcolor{black}{15}\tabularnewline
\hline 
\textcolor{black}{14, 13, $12^{*}$ (mixed) } & \textcolor{black}{15}\tabularnewline
\hline 
\textcolor{black}{11 (2q/p) } & \textcolor{black}{22}\tabularnewline
\hline 
\textcolor{black}{11, 10, 9, 8 (mixed) } & \textcolor{black}{21}\tabularnewline
\hline 
\textcolor{black}{7 (3q/p) } & \textcolor{black}{21}\tabularnewline
\hline 
\end{tabular}
\par\end{centering}
\textcolor{black}{\caption{\label{tab:1}Minimum number of photons and qubits required to reach
our overall information transfer success probability th\textcolor{black}{reshold
of $\overline{P_{S}}=0.995$ with $p_{t}=0.916.$ Similar result}s
are seen for most values of $p_{t}.$ The star corresponds to the
$optimal$ case, in which, by using the mixing strategy, for a given
$N_{\mathrm{min}}$ we reach the lowest $n_{\mathrm{min}}$ for a
specific value of $p_{t}.$ }
}
\end{table}

\textcolor{black}{The number of qubits can be maintained equal to
the non multiplexing case, while reducing the number of photons, with
a $mixed$ strategy, in which each photon can carry an arbitrary number
of qubits (from 1 to 4). Table \ref{tab:1} shows the total number
of photons and the total number of qubits needed for reaching $\overline{P_{S}}$
at $p_{t}=0.916$ using the pure and the mixed strategies. We observe
that we can reach the required $\overline{P_{S}}$ with a lower number
of photons (12) given the same number of qubits (15) when we apply
the mixed strategy. The number of two qubit gates required is therefore
the same as for the non-multiplexing case, even for bigger codes.
This further highlights the potential advantages of quantum multiplexing.
Can these improvements be generalized to other loss based quantum
error correction codes?}

\textcolor{black}{In the quantum Reed\textendash Solomon $[[d,2k-d,d-k+1]]_{d}$
code information is encoded in $d$ qudits, with the code failing
on the loss of $d-k+1$ out of $d$ qudits. For comparative purposes,
we will express the degree of multiplexing as $q$ qubits of information
per photon. When we encode the qudits in these $q$ degrees of freedom
of quantum multiplexed photons, any qudit of information depends upon
the successful transmission of $\lceil\log_{2}(d)/q\rceil$ photons
\cite{Reed_Sal}. The probability of failure is therefore 
\begin{eqnarray}
P_{Fail} & = & \sum_{j=d-k+1}^{d}{d \choose {j}}(1-p_{t}^{\lceil\frac{\log_{2}(d)}{q}\rceil})^{j}p_{t}^{\lceil\frac{\log_{2}(d)}{q}\rceil(d-j)}.
\end{eqnarray}
In this code the block is given by the total number of photons encoding
a single qudit, and if a block is incomplete, the qudit is not successfully
transmitted. Therefore, the performance can be improved by maintaining
independence between these blocks, and by reducing the chances for
loss events within any single block. Adding additional quantum multiplexing
will help so long as it preserves independence between qudit loss
events. For the quantum Reed-Solomon code we can also determine the
lowest number of qubits and photons required to reach $\overline{P_{S}},$
as shown in Fig. \ref{fig:qubits_photons_vs_Pt-1}b. Here, the advantage
of using quantum multiplexed photons is evident in terms of a reduction
of the number of qubits, two qubit gates and photons compared to the
no quantum multiplexing case. In particular, the higher is the quantum
multiplexing degree the less qubits and photons we require. For instance,
at $p_{t}=0.85,$ we have that for $q=4,$ $N_{\mathrm{min}}\simeq40$
and $n_{\mathrm{min}}\simeq10,$ whereas when no quantum multiplexed
photons are in use, we have that both $N_{\mathrm{min}}$ and $n_{\mathrm{min}}$
are over 1000. As $p_{t}$ gets lower, the number of photons and qubits
increases considerably, hence, we need to use higher degrees of quantum
multiplexing. Furthermore, by comparing Fig. (\ref{fig:qubits_photons_vs_Pt-1}a)
with Fig. (\ref{fig:qubits_photons_vs_Pt-1}b), we infer that there
is always a specific value of $q$ for which the Reed-Solomon code
requires a lower number of resources compared to the parity code (for
$q=4,$ at $p_{t}=0.85,$ $N_{\mathrm{min}}(n_{\mathrm{min}})$ is
72\%( 75\%) lower for the Reed-Solomon code than the parity code).
For other error correction codes based on the transmission of qudits
we expect the same reduction in the number of qubits, two qubit gates
and photons when quantum multiplexing is in use.}

\textcolor{black}{There are other loss codes based on encoding information
in superposition of photon number (bosonic \cite{Michael} and GKP
\cite{GKP1} for instance), in which quantum multiplexing is ineffective.
In these cases this would correspond to the assignment of information
about multiple excitation to the various degrees of freedom of a single
mode. However, any quantum multiplexed photon mode is equivalent,
in this case, to a no quantum multiplexed mode. There is always therefore
a code using fewer excitations and a higher number of modes than the
original that will perform as well as the quantum multiplexed case.}

\textcolor{black}{It is essential to compare these quantum multiplexed
codes to the best loss codes currently known - namely the GKP codes
\cite{GKP1}. In Fig. (\ref{fig:qubits_photons_vs_Pt-1}a,b) we plot
(black curves) the average number of photons for the hexagonal GKP
code \cite{gkpref}. This suggests that there are regions where the
GKP code has a better performance and other regions where this is
reversed. The multiplexed codes operate better in the higher loss
regimes. Further, a critical consideration has to be the near deterministic
implementation of the code itself. Our quantum multiplexing approach
requires the same basic two qubit/qudit gates needed for quantum logic
(and the original codes themselves) with the addition of high efficiency
optical switches to swap state between the different degrees of freedom.
On the other hand the GKP code is quite demanding to achieve in a
near deterministic way and necessitates more complex continuous variable
gates. Generating such codes in a heralded but probabilistic fashion
has been achieved but unfortunately increases the resources required
\cite{GKPexp1,GKPexp2} (see Supplemental Material). This indicates
that additional resources will be required at the end nodes to process
the quantum data transmitted over the communication channel.}

\textcolor{black}{To summarize, we have shown how quantum multiplexed
loss codes have the potential to significantly decrease the resources
required to transfer quantum information between two adjacent nodes.
This is achieved while maintaining or even lowering the required number
of two-qubit gates. Two primary error correction codes were considered:
the redundant quantum parity code and the quantum Reed-Solomon code.
For the former, we found that the total number of single photons that
need to be transmitted through the channel can be dramatically reduced
(near 50 percent) without significantly increasing the number of qubits.
Further, we found it advantageous for individual photons to have different
degrees of quantum multiplexing, as well as for blocks to contain
different numbers of qubits. The quantum Reed-Solomon code significantly
outperforms the redundant quantum parity code and, using quantum multiplexed
qudits, has the potential to reduce simultaneously the number of photons,
qubits and gates used. These improvements should be possible in many
(but not all) of the other loss based error correction codes when
quantum multiplexing is used. Quantum multiplexing has the potential
to be a new resource saving tool especially for near term implementations.
Our findings can be applied to any communication system that needs
error correction to improve its efficiency, such as in quantum repeaters,
quantum computation and quantum sensing.}

\textit{\textcolor{black}{We thank Koji Azuma and Peishun Yan for
useful discussions. This project was made possible through the support
of the MEXT KAKENHI Grant-in-Aid for Scientific Research on Innovative
Areas ``Science of Hybrid Quantum Systems\textquotedblright{} Grant
No. 15H05870 and a grant from the John Templeton Foundation (JTF \#
60478). The opinions expressed in this publication are those of the
authors and do not necessarily reflect the views of the John Templeton
Foundation.}}

\bibliographystyle{apsrev4-1}
\bibliography{bib1}

\end{document}